\documentstyle[12pt,epsfig]{article}
\begin{document}
\vspace{-2.0cm} 
\begin{flushright}
TIFR/TH/02-08 \\
hep-lat/0203015 \\
\end{flushright}
\bigskip
\begin{center}
\Large {\bf On The Deconfinement Transition in SU(4) Lattice
Gauge Theory }  \\

\bigskip

\large {Rajiv V. Gavai\footnote {gavai@tifr.res.in}} \\
\bigskip
Department of Theoretical Physics \\
Tata Institute of Fundamental Research\\
Homi Bhabha Road \\
Mumbai 400005, India \\

\bigskip
\bigskip
{\bf ABSTRACT}\\ \end{center}
\bigskip
\noindent  
The deconfinement transition in $SU(4)$ lattice gauge theory is investigated on
$N_s^3 \times N_t$ lattices for $N_s$ = 8--16 and $N_t$ = 4--8 using a modified
Wilson action which is expected to be free of any bulk transitions.  The
susceptibility $\chi_{|L|}^{\rm max}$, where $L$ is the order parameter for
deconfinement, is found to increase {\it linearly} with spatial volume for
$N_t$ = 4, 5, and 6, indicating a first order deconfinement phase transition.
The latent heat of the transition is estimated to be $\approx {2 \over 3}$ of
the corresponding ideal gas energy density at $T_c$.

\newpage

\begin{center}
\bf 1. INTRODUCTION \\
\end{center}
\bigskip

It is widely expected that the ongoing experiments at the Relativistic Heavy
Ion Collider, BNL, New York, and those at the proposed Large Hadron Collider,
CERN, Geneva, will provide us a strong evidence for Quark-Gluon Plasma and may
even yield crucial details about this new phase and the nature of the phase
transition(s) leading to it.  The physics driving this phase transition is
naturally of great interest.  While the real world has two light ($u$,$d$)
flavours of quarks and one somewhat heavier ($s$) flavour, both analytical and
numerical methods in lattice Quantum Chromodynamics (QCD) begin from the
limiting cases of either massless or infinitely massive quarks.  One talks of
the chiral symmetry restoring phase transition and the deconfinement phase
transition in these two cases respectively and has suitable order parameters to
investigate them.  For quarks with $N$ colours and $N_f$ flavours, these
transitions are related to spontaneous breaking of a global $Z(N)$ and $SU(N_f)
\times SU(N_f)$ chiral symmetry, which QCD has in the infinite quark mass and
zero quark mass limit respectively.   Which of them is more relevant in the
real world, is {\it a priori} not clear, since these symmetries are broken
explicitly to various extents.  The low masses of the light flavours suggest
chiral symmetry to be the dominant one, leading one to expect a chiral
transition in QCD at finite temperature.  Indeed, the behaviour of the
corresponding order parameter, the chiral condensate $\langle \bar \psi \psi
\rangle$, confirms this expectation in numerical simulations.  However, it is 
also seen in the same numerical simulations that even the order parameter for
the deconfinement transition, 
\begin{equation}
L(\vec x) = {1 \over N} {\rm tr} \prod_{t=1}^{N_t} U_4(\vec x, t)~~,~~
\nonumber
\end{equation}
acquires nonzero values at this chiral transition; in principle, it could have
done so already at temperatures a lot below the chiral transition since no
symmetry prevents it from doing so.  The energy density also shows a large
change at the transition which would be unexpected from a naive count of
degrees of freedom at a chiral transition but is in fact consistent with that
expected of a deconfinement transition.  These apparently mysterious
observations for QCD with light dynamical quarks can be explained \cite{pt}
using arguments based on the large $N$ limit, {\it if} the deconfinement
transition for $N \ge 4$ is of {\it second order}.  $SU(4)$ is clearly the
simplest case to check whether this is so.

\begin{figure}[htbp]\begin{center}

\epsfig{height=9cm,file=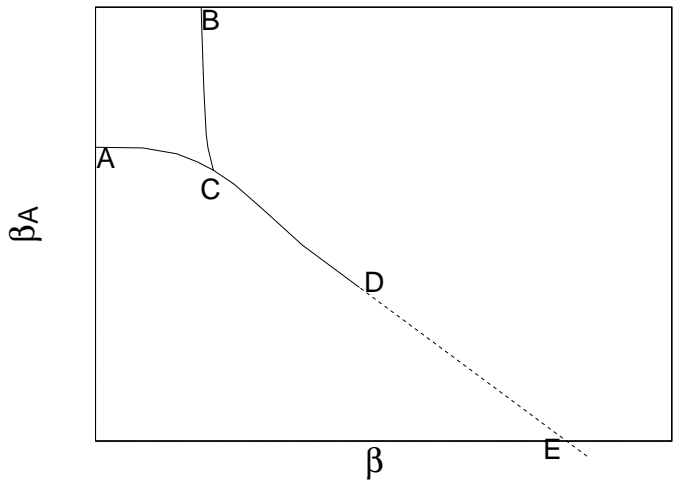, angle=270 }

\caption{A schematic phase diagram in $(\beta, \beta_A)$-plane for the mixed
action of eq. (\protect{\ref{mxd}}).}

\label{fg.phd}\end{center}\end{figure}

Numerical simulations of $SU(4)$ theory at finite temperatures have been
done in the past \cite{old} and recently \cite{recent} as well. All of
them reported a first order deconfinement transition but they all 
used the Wilson action, or the more general mixed action:
\begin{equation}
S = \sum_P \bigg[ \beta \bigg(1 -{{\rm Re~tr}~ U_P\over N}  \bigg) +
    \beta_A  \bigg( 1 - {{\rm tr_A}~ U_P\over N} \bigg) \bigg]~.~
\label{mxd}
\end{equation}
Here $U_P$ denotes the directed product around an elementary plaquette P of the
link (gauge) variables, $U_\mu(x)$, located at site $x$ in the direction $\mu$
=1--4,  $\rm tr $ and $\rm tr_A$ denote the traces of $U_P$ in the fundamental
and adjoint representations, the $\beta$'s are the corresponding couplings,
and $N$ is the number of colours.   The sum over $P$ is over all independent
plaquettes.  A well known problem in the simulations with these actions,
especially for large $N$, is the presence of bulk transitions which are lattice
artifacts.  The generic phase diagram of the mixed action (\ref{mxd}) in its
coupling plane is depicted schematically in Fig.  \ref{fg.phd}.  The solid
lines in it show first order bulk transition lines.  The dotted line after the
end point D is drawn to suggest the impact D may have on the Wilson axis
($\beta_A$ = 0).  For $N \ge 4$, D is expected to be where E is shown, causing
a first order bulk transition for the usual Wilson action.  In order to avoid
it, simulations were made at negative $\beta_A$ \cite{old} and/or for larger
\cite{recent} $N_t$ = 6 and obtained a first order deconfinement phase
transition for $SU(4)$.

Extensive studies \cite{us} of the deconfinement phase transition for the action
above for the $ N = 2$ case have, however, shown that the presence of bulk
transitions affect the order and location of the deconfinement transition in
subtle and somewhat inexplicable ways. For example, recall that one expects the
coupling at which the deconfinement transition takes place to move
logarithmically with $N_t$, the temporal extension of the lattice whereas a
shift in the bulk transition point is expected to be inversely proportional to
the 4-volume of the lattice, $N_s^3N_t$, where $N_s$ is the spatial lattice
size.  This was the rationale behind the choice of simulation parameters in the
earlier \cite{old,recent} studies of the $SU(4)$ theory.    However, we
\cite{us} found that either this expectation is incorrect or there are apparent
qualitative violations of universality for the $SU(2)$ theory since its
deconfinement transition was of second order for small $\beta_A$ but was of
first order for large enough $\beta_A$ for an entire range of $N_t$ from 2
up to 8, with hardly any shift with $N_t$.   Indeed, universality was restored
\cite{us1} in that case only after eliminating the bulk transitions associated
with the $Z(2)$ vortices and $Z(2)$ monopoles by adding large chemical
potentials for them.  It seems natural to expect that the bulk transitions for
$N = 4$ may also have affected the earlier results on the order of its
deconfinement transition for the action above for $N_t$ = 4 or 6.  A cleaner
determination of the order of the deconfinement transition for $SU(4)$ may be
obtained by wiping off the corresponding bulk transitions by suppressing the
relevant $Z(N)$ objects.  We pursue this idea here to investigate the
deconfinement phase transition in the $SU(4)$ theory.

\bigskip
\begin{center}
\bf 2.  SIMULATIONS AND RESULTS \\
\end{center}
\bigskip
Generalizing the idea of positive plaquette models \cite{pp} in the 
literature for the $SU(2)$ lattice gauge theory to the case of $SU(N)$
theories, we propose to simulate the action
\begin{equation}
S = \beta \sum_P \bigg( 1 - {{\rm Re~tr}~ U_P \over N}\bigg) \cdot \theta
\bigg( {\pi \over N} -|\alpha| \bigg) ~,~
\label{ppm}
\end{equation}
for $N=4$, where $ -\pi < \alpha \le \pi$ is the phase of tr $U_P$.  By adding
the adjoint term of eq. (2) to the action (3), one sees that the phase diagram
of the resultant mixed action should not have the bulk lines AC or BC and hence
the endpoint D or E.  This is because the action (\ref{ppm}) restricts 
${\rm tr}~U_P$ to lie only in one $Z(N)$-sector for all, even small, $\beta$.

We have simulated the above action on $N_s^3 \times N_t$ lattices for
$N_s$ =8, 10, 12, 15, 16 and $N_t$ = 4, 5, 6, 8 using a 15-hit Metropolis et al.
algorithm. The calculations were done on a cluster of pentiums.
Typically short runs to look for points of rapid variations in
$ \langle |L| \rangle$ were followed by long runs (a few million sweeps)
to determine the susceptibility $\chi_{|L|}$,
\begin{equation}
\chi_{|L|}~ =N_s^3~(\langle |L|^2 \rangle - \langle |L| \rangle ^2) ~,~
\label{chiL}
\end{equation}
as a function of $\beta$ using the histogramming technique \cite{FerSwe}.  In
order to monitor the presence of any bulk transition, we also studied the
plaquette susceptibility given by,
\begin{equation}
\chi_P~ =6 N_s^3 N_t~(\langle P^2 \rangle - \langle P \rangle^2)~,~
\label{chiB}
\end{equation}
where P denotes the average of (Re tr $U_P$)/N over the entire lattice.  Usual
finite size scaling techniques were used to determine the order of the
transition and its exponents. According to the finite size scaling
theory \cite{Barb}, the peak of the $|L|$ (or plaquette) susceptibility at the
location of the deconfinement (or bulk) transition should grow on $N_s^3
\times N_t$ lattices like
\begin{equation}
\chi^{\rm max}_{|L|~{\rm or}~P} \propto N_s^{\omega}~,~
\label{chifs}
\end{equation}
for fixed $N_t$.  For a second order transition, $\omega \equiv \gamma/\nu$,
where $\gamma$ and $\nu$ characterize the growth of the $|L|$
(plaquette)-susceptibility and the correlation length near the critical
temperature (coupling) on an infinite spatial lattice.  If the phase transition
were to be of first order instead, then one expects \cite{ChLaBi} the exponent
$\omega = 3$, corresponding to the dimensionality of the space.  In addition,
of course, the $\langle |L| \rangle $ or $\langle P \rangle$ is expected to
exhibit a sharp, or even discontinuous, jump. The corresponding probability
distribution should then show a double peak structure in stead of a usual
gaussian distribution.

\begin{figure}[htbp]\begin{center}
\epsfig{height=13cm,width=14cm,file=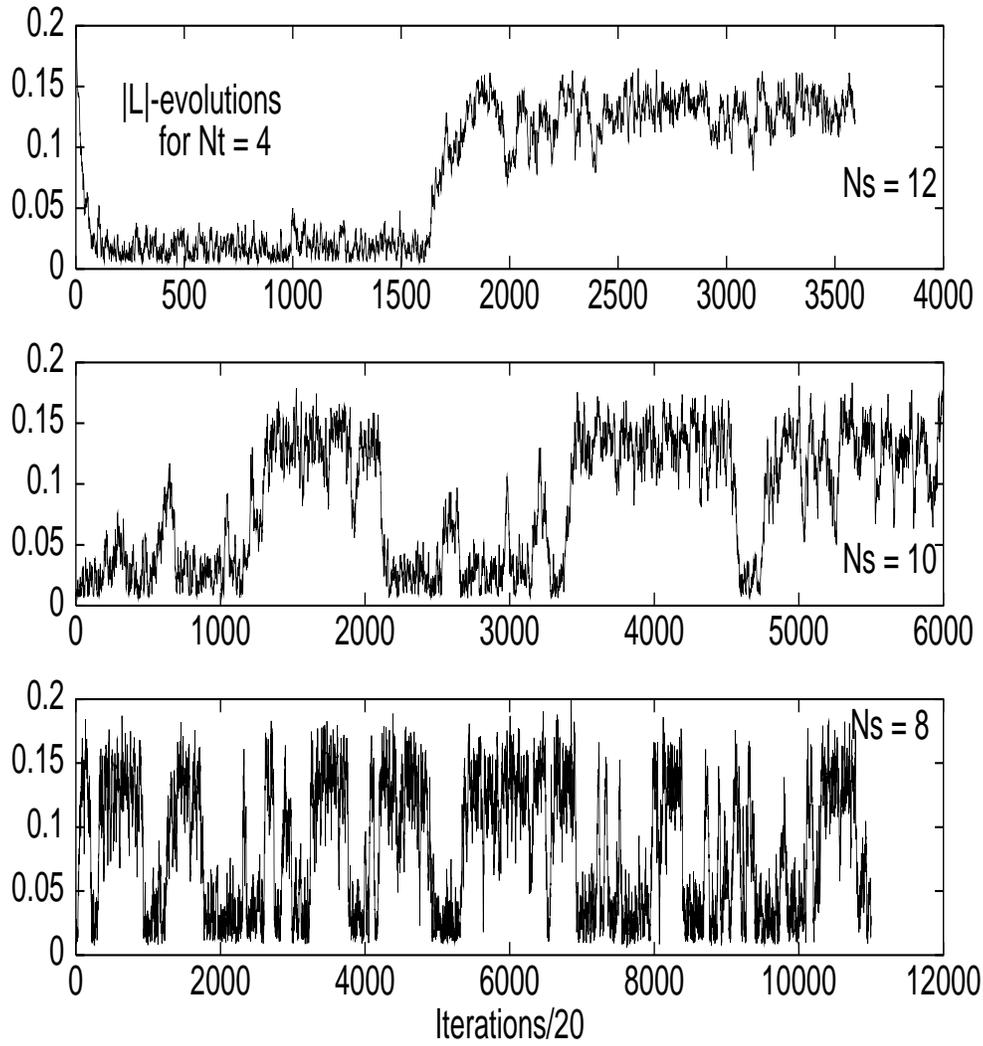,angle=270}
\caption{ The evolution of $|L|$ as a function of Monte Carlo time for
$N_s^3 \times 4$ lattices close to the transition point. }
\label{fg.evol}\end{center}\end{figure}

\begin{figure}[htbp]\begin{center}
\epsfig{height=11cm,width=7cm,file=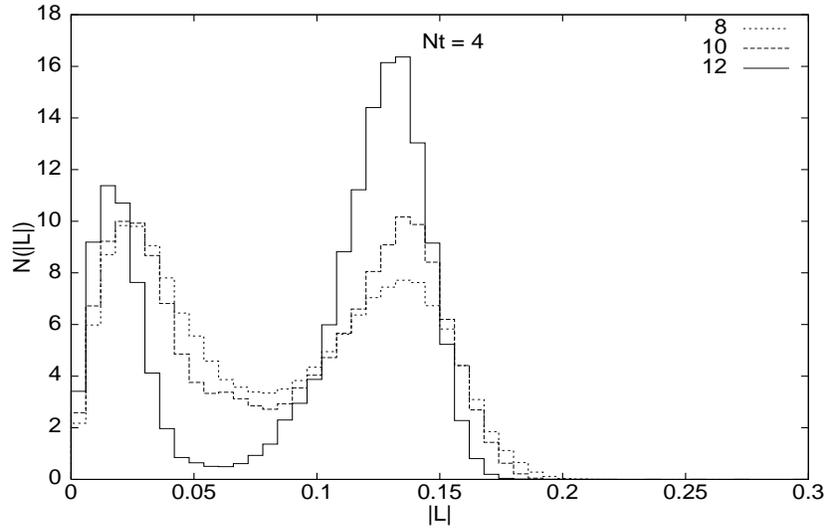,angle=270}
\caption{ The histograms of $|L|$ on $N_s^3 \times 4$ lattices
for long runs close to the transition point. }
\label{fg.hisl}\end{center}\end{figure}

\begin{figure}[htbp]\begin{center}
\epsfig{height=11cm,width=7cm,file=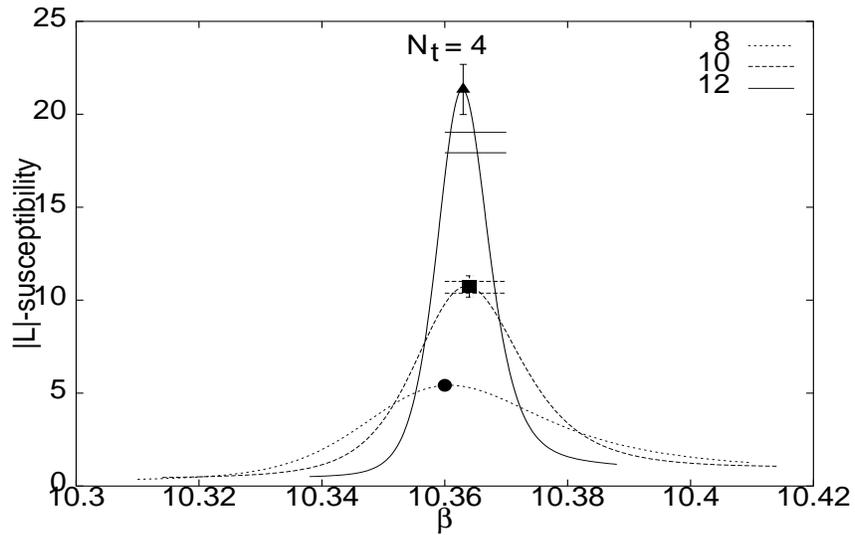,angle=270}
\caption{ The susceptibility $\chi_{|L|}$ as a function of $\beta$ for
$N_s^3 \times 4$ lattices. }
\label{fg.susc}\end{center}\end{figure}

\begin{figure}[htbp]\begin{center}
\epsfig{height=11cm,width=7cm,file=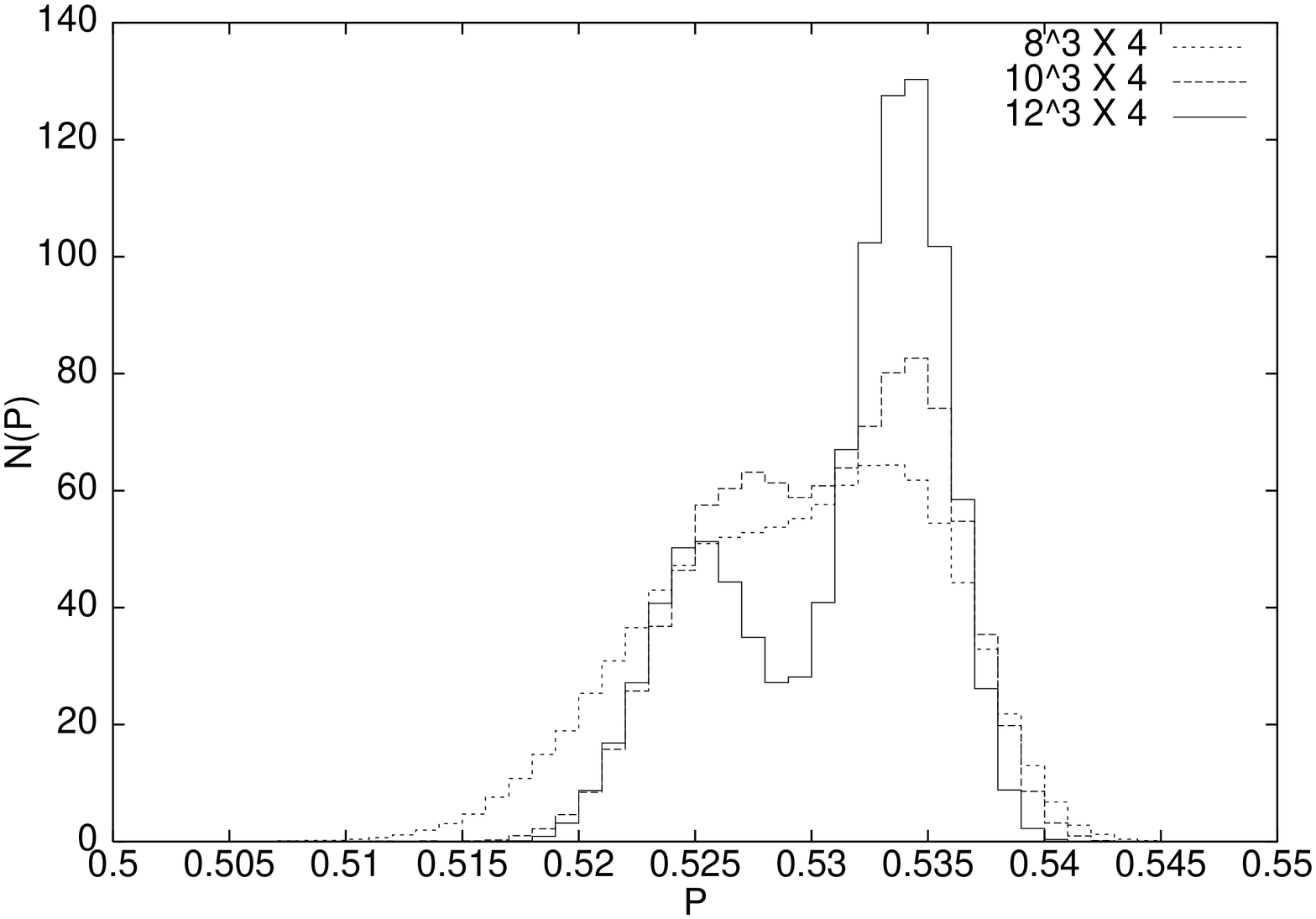,angle=270}
\caption{ The histograms of the plaquette $P$ on $N_s^3 \times 4$ lattices
for the same runs as in Fig. \ref{fg.hisl}. }
\label{fg.hisp}\end{center}\end{figure}

In simulations on $N_s^3 \times 4$ lattices, $N_s =$ 8, 10, 12, we found the
hot (all link variables random) and cold (all links equal to identity) starts 
to converge quickly to a unique value of $|L|$
at couplings a little away from the transition point on its
both sides but a clear co-existence of states was visible for all lattices at
the transition point, as shown in Fig. \ref{fg.evol}. It is clearly seen that
the tunnelling frequency goes down with spatial volume.  The histograms of 
$|L|$, shown in Fig. \ref{fg.hisl}, display peaks which become narrower with 
increasing volume while the gap between them remains unchanged.  These classic 
signs of a first order phase transition are confirmed by a quantitative 
analysis of the growth of $\chi_{|L|}$ with volume, as seen in 
Fig. \ref{fg.susc}.   The horizontal lines in each case are predictions 
obtained by scaling the $N_s = 8$ results linearly with volume, as expected
from eq. (\ref{chifs}) and $\omega=3$ for a first order transition.

Although the above first order transition is clearly a deconfinement
transition, as the behaviour of the order parameter $|L|$ certifies, it could,
of course be due to a coincident bulk transition.  One can check for this
possibility by analogous studies of the average plaquette $P$.  Fig.
\ref{fg.hisp} displays the histograms of $P$ for the same runs as in Figs.
\ref{fg.evol} and \ref{fg.hisl}.  These histograms do exhibit curious
non-gaussian structures, with a suggestion to develop a discontinuity as the
spatial volume grows. The corresponding susceptibilities turn out to be too
noisy to be conclusive, especially when compared to the behaviour of
$|L|$-susceptibility in Fig. \ref{fg.susc}.  However, their growth with volume 
is not inconsistent with being linear.  We therefore made simulations on larger
symmetric lattices up to the sizes $12^4$ at these couplings and in their
immediate neighbourhood and found only usual gaussian distributions for the
plaquette.  For a genuine bulk transition, the behaviour in Fig.  \ref{fg.hisp}
should have been accentuated with the three-fold increase in the 4-volume.
Recall that the presence of latent heat at a first order deconfinement phase
transition can also be a source of the behaviour in Fig. \ref{fg.hisp}. This
should then lead to specific predictions as the temporal lattice size is
increased.  We postpone discussing them in the next section, turning in stead
to investigations of larger $N_t$.

\begin{figure}[htbp]\begin{center}
\epsfig{height=11cm,width=7cm,file=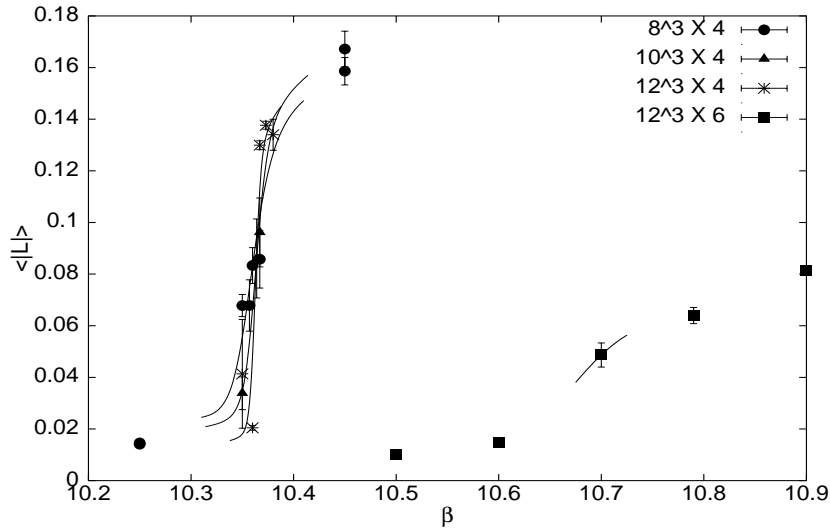,angle=270}
\caption{ $|L|$ as a function of $\beta$ for
lattices with $N_t$ 4 and 6. }
\label{fg.bothnt}\end{center}\end{figure}

For larger $N_t$, we used many longer runs in the region of strong variation of
$\langle |L| \rangle$ to obtain the susceptibility directly and used the
histogramming technique only for the finer determinations of the critical
coupling.  Our results for $\langle |L| \rangle$ as a function of $\beta$
clearly show the expected shift for a deconfinement phase transition for $N_t$
= 4 and 6, as can be seen in Fig. \ref{fg.bothnt}.  
As is usual for the $SU(2)$ and $SU(3)$ theories, the rise of the order
parameter at $\beta_c$ is seen to be slower in Fig. \ref{fg.bothnt} as $N_t$ is
increased due to the well-known ultra-violet effects.  The $\beta_c$ values,
determined from the $\chi_{|L|}^{\rm max}$ and listed in Table \ref{table:1}
for $N_t$ = 4, 5 and 6 for different spatial volumes, also evidently suggest
strong shifts in $\beta_c$ with $N_t$.  Comparing our results in Table
\ref{table:1} with those of Ref. \cite{recent} for $N_t=6$ for the usual Wilson
action, i.e, for eq.(\ref{mxd}) with $\beta_A$ = 0, one finds a shift by
$\approx 0.1$ towards smaller coupling.  Using in each case the peak height for
the smaller spatial volume, the $\chi^{\rm max}$ on the bigger lattice can be
predicted, assuming a first order deconfinement phase transition.   The
predictions listed in Table \ref{table:1} can be seen to be in very good
agreement with the direct Monte Carlo determinations.  Along with the shifts in
$\beta$, these confirm that the same {\it physical} first order deconfinement
phase transition is being simulated on these lattices as the continuum limit of
$a \to 0$ is approached in a progressive manner by increasing the temporal
lattice size $N_t$.

\begin{table}
\caption{The values of $\beta$ at which long simulations were performed 
on $N_s^3 \times N_t$ lattices, $\beta_c$ and the height of 
the $|L|$-susceptibility peak, $\chi^{\rm max}_{|L|}$.}
\label{table:1}
\medskip
\begin{center}
\begin{tabular}{@{}lllll}
\hline
$N_s^3 \times N_t $ & $\beta$  &  $\beta_{c,N_s}$ & $\chi_{|L|}^{\rm max}$& $\chi^{\rm max}_{\rm predicted}$ \\
\hline
$  8^3 \times 4 $     &  10.360   &  10.360(2)     &  5.48(16) &   --      \\
$ 10^3 \times 4 $     &  10.364   &  10.364(5)     &  11.16(35) & 10.71(32) \\
$ 12^3 \times 4 $     &  10.363   &  10.363(5) &  22.41(1.11)&  18.5(6)  \\
\hline
$ 10^3 \times 5 $     &  10.520   &  10.515(5)     &  4.63(40) &   --      \\
$ 15^3 \times 5 $     &  10.520   &  10.525(5) &  14.02(1.50)&  15.6(1.3)  \\
\hline
$ 12^3 \times 6 $     &  10.675   &  10.686(5)     &  4.36(35) &   --      \\
$ 16^3 \times 6 $     &  10.675   &  10.676(5) &  10.43(95)&  10.3(8)  \\
\hline
\end{tabular} 
\end{center}
\
\end{table}

\bigskip 

\begin{center}
\bf 3. SCALING AND LATENT HEAT\\
\end{center}

\bigskip

\begin{figure}[htbp]\begin{center}
\epsfig{height=11cm,width=7cm,file=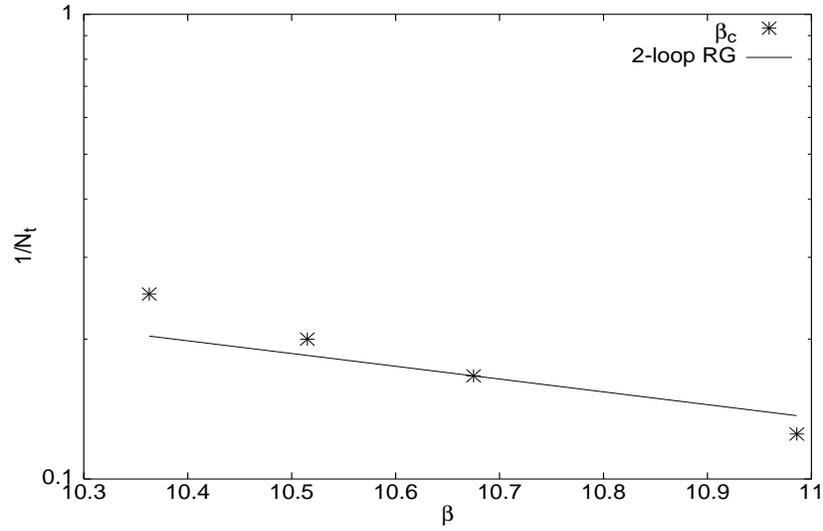,angle=270}
\caption{ $N_t^{-1}$ as a function of $\beta$ along with the 
RG-curve mentioned in the text.}
\label{fg.cont}\end{center}\end{figure}

\begin{figure}[htbp]\begin{center}
\epsfig{height=11cm,width=7cm,file=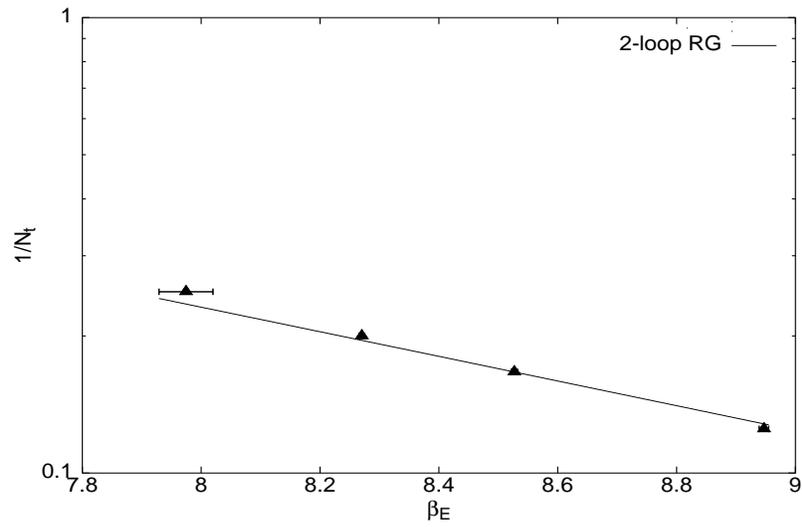,angle=270}
\caption{ $N_t^{-1}$ as a function of $\beta_E$ along with the corresponding
RG-curve.}
\label{fg.contp}\end{center}\end{figure}

A quantitative test of the fact that the first order transition reported in the
previous section is indeed a physical deconfinement transition (and not a bulk
transition) consists of translating all the $\beta_c$ in Table \ref{table:1} to
the corresponding transition temperature in physical units and thus checking
whether it is constant while the $\beta_c$'s shift with $N_t$ as expected from
the renormalization group equation,

\begin{equation}
 aT_c = {1 \over N_\tau} = {T_c \over \Lambda_L} \left({4b_0 \over \beta }\right)^{-b_1/b_0^2} 
\exp \left(- {\beta \over 8b_0}\right)~~,~~
\label{2lprg}
\end{equation}
where 
\begin{equation}
b_0 = {11 \over 12 \pi^2}, ~~{\rm and}~~ b_1 = {17 \over 24 \pi^4}~~,~~
\label{b0b1}
\end{equation}
are the first two coefficients of the perturbative $\beta$-function for
the $SU(4)$ Yang-Mills theory.  Fig. \ref{fg.cont} exhibits a comparison of
our set of $\beta_c$ for $N_t$ =4, 5, 6 and 8 with eq.(\ref{2lprg}).  The
normalization was chosen by demanding the line to pass through the point at
$N_t=6$. While the shifts are in accord with the expectations, one also sees
strong quantitative deviations.  These are, however, not unusual: similar
deviations have been seen in the studies for $N$ = 2 and 3 \cite{Urs2} as 
well.   As pointed earlier \cite{parisi}, and tested successfully \cite{Urs2}
for $SU(2)$ and $SU(3)$, one possible cure for removing
these deviations is to employ a better choice of the coupling in 
eq.(\ref{2lprg}) which consists of replacing $\beta$ there by 
\begin{equation}
\beta_E = { 0.25(N^2 -1) \over {1 - \langle P \rangle }}~.~
\label{parb}
\end{equation}
Fig. \ref{fg.contp} displays our data using this new variable along with
the corresponding RG-curve of eq.(\ref{2lprg}) normalized the same way as 
before.  One sees an excellent agreement, providing a concrete quantitative
evidence in favour of the physical nature of the transition.   Furthermore,
it leads to an estimate of $T_c$ = $ 18.5 \pm 0.5 \Lambda_L$, where the 
error is estimated by requiring all the data points to lie within the
band generated by it.

\begin{figure}[htbp]\begin{center}
\epsfig{height=11cm,width=7cm,file=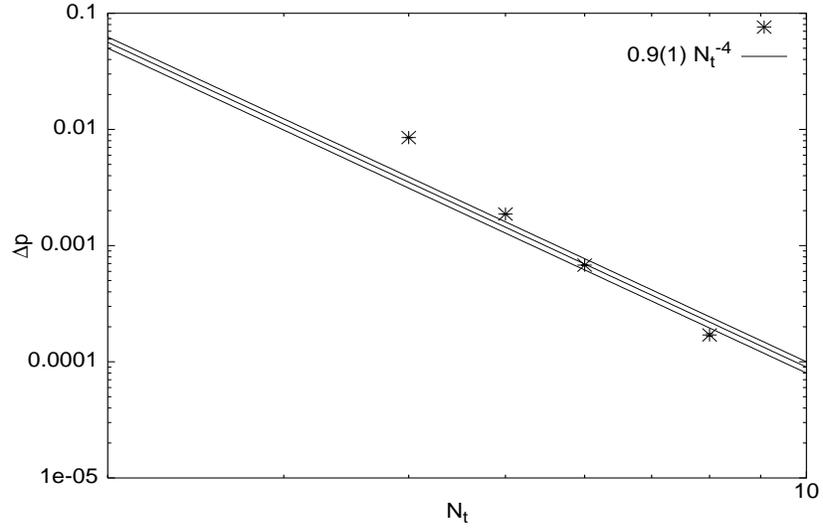,angle=270}
\caption{ The plaquette discontinuity $\Delta P $ as a function 
of $N_t$.  }
\label{fg.plad}\end{center}\end{figure}

\begin{figure}[htbp]\begin{center}
\epsfig{height=11cm,width=7cm,file=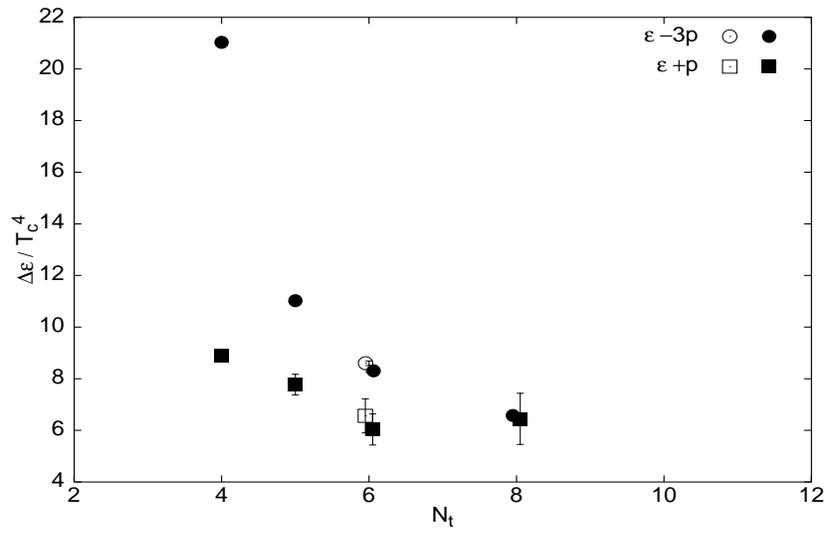,angle=270}
\caption{ The two latent heat heat estimates of eq. (\ref{lat}) as a function 
of $N_t$.  }
\label{fg.laheat}\end{center}\end{figure}

Apart from the characteristic (logarithmic) shift of the transition point with
$N_t$, seen above, the latent heat of a first order deconfinement phase
transition provides yet another quantitative test of the continuum limit since
it should also remain constant as $N_t \to \infty$.  Requiring the pressure to
be continuous at the deconfinement phase transition, the latent heat can be
obtained from two different observables $ \Delta_1 \equiv \Delta
(\epsilon-3p)/T_c^4$, and $ \Delta_2 \equiv \Delta (\epsilon+p)/T_c^4$, 
where $\Delta$ denotes discontinuities across the transition in the 
respective variables,  
\begin{eqnarray}
&& \Delta_1  = - 48 N_t^4 a {\partial g^{-2} \over \partial a }
\Delta P \nonumber~,~ \\
&& \Delta_2  = 32 N_t^4 {C(g^2)\over g^2} (\Delta P_t -\Delta
P_s)~,~ 
\label{lat}
\end{eqnarray}
and $C(g^2) = (1 - 0.2366 g^2 + O(g^4))$ for $SU(4)$ \cite{recent}. We will
employ the perturbative $\beta$-function in $\Delta_1$ with its coefficients
given by eq.  (\ref{b0b1}).  In order to obtain the $\Delta P$, $\Delta P_s$
and $\Delta P_t$, the minimum of the histogram $N(|L|)$ was used to separate
the two phases in each case.  The errors were estimated by varying it within
the limits suggested by the corresponding histograms.  From eq. (\ref{lat}), it
is clear that the plaquette discontinuity $\Delta P \propto N_t^{-4}$ if one is
to obtain the same latent heat in physical units, since the rest of the factors
vary mildly with $N_t$.  In case of a bulk phase transition, on the other hand,
one would expect the plaquette discontinuity to remain constant or exhibit a
mild increase.  Indeed, as Fig.  \ref{fg.plad} displays, its decrease with
$N_t$ is seen to be consistent with these expectations for $N_t \ge 5$.  Our
estimates of latent heat from $\Delta_1$ and $\Delta_2$ are shown in Fig.
\ref{fg.laheat} as a function of $N_t$ (filled symbols) along with the
corresponding results of Ref.  \cite{recent} for $N_t$ = 6 (open symbols) which
can be seen to be in good agreement with ours.  Both the estimates in
eq.(\ref{lat}) must agree with each other as $N_t \to \infty$, since the
neglected cut-off corrections become then insignificant.  Table \ref{table:2}
and Fig.  \ref{fg.laheat} verify this to be the case.  One may therefore take
the results for $N_t$ =8 as a reasonable approximation to the continuum limit.
Comparing them with the energy density of the corresponding ideal gas at $T_c$,
$\epsilon_{SB} = \pi^2 (N^2 -1) T_c^4/15 = \pi^2 T_c^4$, one obtains for the
latent heat,
\begin{equation}
{\Delta \epsilon \over \epsilon_{SB}} = 0.6657 \pm 0.0162~.~
\label{lasb}
\end{equation}
Here we have used $\Delta_1$ to get the result above. Using $\Delta_2$ leads
to $0.65 \pm 0.10$.

\begin{table}
\caption{Both the latent heat estimates of eq.(\ref{lat}) as a function of
$N_t$.}
\label{table:2}
\medskip
\begin{center}
\begin{tabular}{@{}lllll}
\hline
$N_t$ & 4 & 5 & 6 & 8  \\ 
\hline
$\Delta_1 $&  21.03(5) &  11.02(6) &  8.31(5) &  6.57(16) \\
$\Delta_2$ &  9.89(14) &  7.77(40) &  6.04(60) &  6.45(99)\\
\hline
\end{tabular} 
\end{center}
\
\end{table}

\newpage

\begin{center}
4.  \bf SUMMARY AND DISCUSSION \\
\end{center}
\bigskip

Establishing the order of the deconfinement phase transition for the $SU(4)$
lattice gauge theory at finite temperature is important for our understanding
of the physics of the phase transition to quark-gluon plasma.  Possible subtle
influences, which the bulk transitions for the mixed action of eq. (\ref{mxd})
(see Fig. \ref{fg.phd}) may have, necessitate an approach to bypass them.  In
this paper, we have attempted this by generalizing the idea of positive
plaquette model for the $SU(2)$ case and simulating the action of eq.
(\ref{ppm}) to investigate the $SU(4)$ theory at $ T \ne 0$ on $N_s^3 \times
N_t$ lattice with $N_s \ge 2N_t$ and $N_t$ = 4, 5, 6 and 8.  

Various qualitative indicators, such as, the histograms and evolutions of the
order parameter $L$, suggest a {\it first} order deconfinement phase transition
for $SU(4)$ on $N_t=4$ lattices with $N_s$ varying up to 12.  The linear
growth of $\chi_{|L|}^{\rm max}$ with volume for $N_t$ = 4, quantitatively
confirms this finding.  Increasing $N_t$ to 5 and 6, one again finds a 
growth in $\chi_{|L|}^{\rm max}$ that is consistent with being linear in
volume. Defining the transition coupling as the location at which the 
maximum of the above susceptibility occurs, one finds a significant shift
in it as $N_t$ is varied from 4 to 8. The amount of shift is 
consistent with the expectations of a physical deconfinement transition;
a bulk transition would have had much smaller shifts.  Indeed, the set
of $\beta_c$ for $N_t$ =4, 5, 6 and 8 is consistent with the asymptotic 
scaling relation of eq.(\ref{2lprg}) when used in conjunction with the 
improved coupling of eq. (\ref{parb}), leading to a determination of the 
transition temperature $T_c = (18.5 \pm 0.5) \Lambda_L$.

The presence of latent heat at a first order phase transition necessarily means
a discontinuity in the average plaquette on a finite lattice, which can however
also be due to an unphysical bulk transition.  We find that the plaquette
discontinuity $\Delta P$ decreases with increase in $N_t$ as approximately the
fourth power of $N_t$. This indicates both a lack of a bulk transition and the
presence of a first order deconfinement phase transition, since it suggests i)
a vanishing discontinuity in the $N_t \to \infty$ limit ii) but in a manner
that leaves the latent heat constant in physical units.  We estimate the latent
heat to be $(0.666 \pm 0.016)$ times the corresponding ideal gas energy density
at $T_c$ using the perturbative $\beta$-function in eq.  (\ref{lat}).
This value, obtained from our $N_t$ =8 simulation, is expected to be close to
the continuum limit since the two different estimates from eq. (\ref{lat})
coincide.  Comparing with the corresponding value (using again perturbative
$\beta$-function) for the $SU(3)$ theory, which is about 0.43 \cite{bkp}, one
finds that the deconfinement transition grows stronger in nature as the number
of colours $N$ is increased.  Whether this continues as $N$ is increased
further is at present not clear.  However, already the known results for $N$ =
3 and 4 are sufficient to cast doubt on the explanation proposed in Ref.
\cite{pt}.  Consequently, the mystery of the behaviour of the energy density
and the deconfinement order parameter, $L$, mentioned in the introduction 
remains.

\bigskip 

\begin{center}
5.  \bf ACKNOWLEDGMENTS \\
\end{center}
\bigskip 
It is a pleasure to acknowledge interesting discussions with Sourendu
Gupta.

\end{document}